\documentclass[12pt]{article}

\usepackage{mathrsfs}
\usepackage[T1]{fontenc}
\usepackage{mathpazo}
\usepackage{setspace}
\usepackage{amsfonts}
\usepackage{amssymb}
\usepackage{epsfig}
\usepackage{latexsym}
\usepackage{color}
\usepackage{graphicx}
\usepackage{nicefrac}
\usepackage[latin1]{inputenc}
\usepackage{pstricks}
\usepackage{slashed}
\usepackage{multirow}

\def\hybrid{\topmargin -30pt    \oddsidemargin 0pt 
        \headheight 0pt \headsep 0pt
        \textwidth 6.25in       
        \textheight 9.5in       
        \marginparwidth .875in
        \parskip 5pt plus 1pt   \jot = 1.5ex}

\hybrid

\def\baselinestretch{1.2}

\catcode`\@=11

\def\marginnote#1{}
\newcount\hour
\newcount\minute
\newtoks\amorpm
\hour=\time\divide\hour by60
\minute=\time{\multiply\hour by60 \global\advance\minute by-\hour}
\edef\standardtime{{\ifnum\hour<12 \global\amorpm={am}%
        \else\global\amorpm={pm}\advance\hour by-12 \fi
        \ifnum\hour=0 \hour=12 \fi
        \number\hour:\ifnum\minute<10 0\fi\number\minute\the\amorpm}}
\edef\militarytime{\number\hour:\ifnum\minute<10 0\fi\number\minute}

\def\draftlabel#1{{\@bsphack\if@filesw {\let\thepage\relax
   \xdef\@gtempa{\write\@auxout{\string
      \newlabel{#1}{{\@currentlabel}{\thepage}}}}}\@gtempa
   \if@nobreak \ifvmode\nobreak\fi\fi\fi\@esphack}
        \gdef\@eqnlabel{#1}}
\def\@eqnlabel{}
\def\@vacuum{}
\def\draftmarginnote#1{\marginpar{\raggedright\scriptsize\tt#1}}

\def\draft{\oddsidemargin -.5truein
        \def\@oddfoot{\sl preliminary draft \hfil
        \rm\thepage\hfil\sl\today\quad\militarytime}
        \let\@evenfoot\@oddfoot \overfullrule 3pt
        \let\label=\draftlabel
        \let\marginnote=\draftmarginnote
   \def\@eqnnum{(\theequation)\rlap{\kern\marginparsep\tt\@eqnlabel}%
\global\let\@eqnlabel\@vacuum}  }

\def\draft2{
        \def\@oddfoot{\sl preliminary draft \hfil
        \rm\thepage\hfil\sl\today\quad\militarytime}
        \let\@evenfoot\@oddfoot \overfullrule 3pt
        \let\label=\draftlabel
        \let\marginnote=\draftmarginnote
   \def\@eqnnum{(\theequation)\rlap{\kern\marginparsep\tt\@eqnlabel}%
\global\let\@eqnlabel\@vacuum}  }

\def\preprint{\twocolumn\sloppy\flushbottom\parindent 2em
        \leftmargini 2em\leftmarginv .5em\leftmarginvi .5em
        \oddsidemargin -.5in    \evensidemargin -.5in
        \columnsep .4in \footheight 0pt
        \textwidth 10.in        \topmargin  -.4in
        \headheight 12pt \topskip .4in
        \textheight 6.9in \footskip 0pt
        \def\@oddhead{\thepage\hfil\addtocounter{page}{1}\thepage}
        \let\@evenhead\@oddhead \def\@oddfoot{} \def\@evenfoot{} }

\def\numberbysection{\@addtoreset{equation}{section}
        \def\theequation{\thesection.\arabic{equation}}}

\def\underline#1{\relax\ifmmode\@@underline#1\else
        $\@@underline{\hbox{#1}}$\relax\fi}

\def\titlepage{\@restonecolfalse\if@twocolumn\@restonecoltrue\onecolumn
     \else \newpage \fi \thispagestyle{empty}\c@page\z@
        \def\thefootnote{\fnsymbol{footnote}} }

\def\endtitlepage{\if@restonecol\twocolumn \else \newpage \fi
        \def\thefootnote{\arabic{footnote}}
        \setcounter{footnote}{0}}  

\catcode`@=12
\relax

\def\figcap{\section*{Figure Captions\markboth
        {FIGURECAPTIONS}{FIGURECAPTIONS}}\list
        {Figure \arabic{enumi}:\hfill}{\settowidth\labelwidth{Figure
999:}
        \leftmargin\labelwidth
        \advance\leftmargin\labelsep\usecounter{enumi}}}
 \relax
\def\tablecap{\section*{Table Captions\markboth
        {TABLECAPTIONS}{TABLECAPTIONS}}\list
        {Table \arabic{enumi}:\hfill}{\settowidth\labelwidth{Table
999:}
        \leftmargin\labelwidth
        \advance\leftmargin\labelsep\usecounter{enumi}}}
 \relax
\def\reflist{\section*{References\markboth
        {REFLIST}{REFLIST}}\list
        {[\arabic{enumi}]\hfill}{\settowidth\labelwidth{[999]}
        \leftmargin\labelwidth
        \advance\leftmargin\labelsep\usecounter{enumi}}}
 \relax
\makeatletter
\newcounter{pubctr}
\def\publist{\@ifnextchar[{\@publist}{\@@publist}}
\def\@publist[#1]{\list
        {[\arabic{pubctr}]\hfill}{\settowidth\labelwidth{[999]}
        \leftmargin\labelwidth
        \advance\leftmargin\labelsep
        \@nmbrlisttrue\def\@listctr{pubctr}
        \setcounter{pubctr}{#1}\addtocounter{pubctr}{-1}}}
\def\@@publist{\list
        {[\arabic{pubctr}]\hfill}{\settowidth\labelwidth{[999]}
        \leftmargin\labelwidth
        \advance\leftmargin\labelsep
        \@nmbrlisttrue\def\@listctr{pubctr}}}
 \relax
\makeatother

\def\be{\begin{equation}}
\def\ee{\end{equation}}
\def\ba{\begin{eqnarray}}
\def\ea{\end{eqnarray}}

\def\del{\partial}

\def\a{\alpha}

\def\b{\beta}

\def\g{\gamma}
\def\G{\Gamma}
\def\d{\delta}

\def\e{\epsilon}

\def\th{\theta}

\def\m{\mu}
\def\n{\nu}
\def\om{\omega}
\def\Om{\Omega}
\def\l{\lambda}

\def\s{\sigma}

\def\cL{{\cal L}}

\def\no{\noindent}

\def\qq{\qquad}

\def\IR{\relax{\rm I\kern-.18em R}}

\def\inv{^{\raise.0ex\hbox{${\scriptscriptstyle -}$}\kern-.05em 1}}

\def \ha {{\frac{1}{2}}}

\def \ov {\over}

\begin{document}


\renewcommand{\theequation}{\thesection.\arabic{equation}}
\csname @addtoreset\endcsname{equation}{section}

\begin{titlepage}
\begin{center}

{}\hfill QMUL-PH-10-08

\phantom{xx}
\vskip 0.5in

{\large \bf  Canonical pure spinor (Fermionic) T-duality }

\vskip 0.5in

{\bf K. Sfetsos}${}^{1a}$,\phantom{x}
{\bf K. Siampos}${}^{1b}$\phantom{x}and\phantom{x}{\bf Daniel C. Thompson}${}^{2c}$
\vskip 0.1in

${}^1$Department of Engineering Sciences, University of Patras,\\
26110 Patras, Greece\\

\vskip .2in

${}^2$ Queen Mary University of London, Centre for Research in String Theory, \\
Department of Physics, Mile End Road,
London, E1 4NS, United Kingdom\\

\end{center}

\vskip .4in

\centerline{\bf Abstract}

\no
We establish that the recently discovered fermionic T-duality
can be viewed as a canonical transformation in phase space. This
requires a careful treatment of constrained Hamiltonian systems.
Additionally, we show how the canonical
transformation approach for bosonic T-duality can be extended to include Ramond--Ramond
backgrounds in the pure spinor formalism.

\vfill
\no
 {
 $^a$sfetsos@upatras.gr,\phantom{x}
 $^b$siampos@upatras.gr,\phantom{x} $^c$d.c.thompson@qmul.ac.uk}

\end{titlepage}
\vfill
\eject



\def\baselinestretch{1.2}
\baselineskip 20 pt
\no

\newcommand{\eqn}[1]{(\ref{#1})}
\section{Introduction}

An important recent development in the study of ${\cal N}=4$
supersymmetric gauge theories has been the discovery of a connection
between planar scattering amplitudes and Wilson loops and the
related discovery of a dual superconformal symmetry. From the dual
$AdS$ perspective this result is understood (at least at strong
coupling) as a consequence of T-duality
\cite{Alday:2007hr,Berkovits:2008ic,Beisert:2008iq,Ricci:2007eq}.
 In  \cite{Alday:2007hr} it was established that under a series of T-dualities the $AdS_5\times S_5$
 metric is self-dual and moreover, that a configuration corresponding to a scattering amplitude
 is dualised to one describing a light-like Wilson loop.

\no
However, the dualities used in \cite{Alday:2007hr}  do not
leave the full $AdS$ background strictly invariant;  instead they
result in a shifted dilaton and different Ramond--Ramond (RR) fields.
  To rectify this and produce an exact selfduality of the background
   Berkovits and Maldacena  \cite{Berkovits:2008ic} introduced
   a novel  ``Fermionic T-duality''  which  leaves the metric and
    Kalb--Ramond fields invariant but transforms the dilaton and RR fields.
     Whilst it is only valid at tree level in string perturbation theory
     and not a full symmetry of string theory, Fermionic T-duality is
     clearly important and certainly has applications as a solution
     generating symmetry of supergravity \cite{Bakhmatov:2009be}.

\no
The derivation of Fermionic T-duality in  \cite{Berkovits:2008ic}
 essentially follows the Buscher procedure carried out along the
 direction of a fermionic isometry in superspace.
It has long been known that an alternative way to think about T-duality,
albeit classical in nature, is as a canonical transformation
of the phase space variables. This was first shown in the context of the chiral $O(4)$ bosonic model
(dualized using its non-abelian symmetry) in \cite{Curtright:1994be} and for
abelian T-duality of the bosonic string in \cite{Alvarez:1994wj}. It was
later extended to the RNS formalism
of the superstring \cite{Hassan:1995je} and also to the more general notions of non-abelian and
 Poisson Lie T-duality in \cite{Lozano:1995jx} and \cite{Sfetsos:1997pi}, respectively.

\no
 In this letter we will show how this canonical approach may be
 extended to Fermionic T-duality.   Because we are dealing with fermions
 we will see that proving the canonical equivalence requires a careful
 application of the Dirac procedure in order to treat second class constraints.
   As a by-product of this study we shall also show that the canonical
    transformation approach for bosonic T-duality can be readily
    extended to the pure spinor form of the superstring thus incorporating
    the transformations of both NS, RR and fermionic background fields.

\no
We believe that our work will be useful in extending the notion of T-duality in
superstring theory in the presence of non-trivial RR background fields when non-abelian
isometry structures are involved.

\section{Bosonic T-duality as a canonical transformation}

For later reference we begin by reviewing the canonical transformation approach to plain bosonic T-duality. We shall see that certain structures are the same even when many more fields are present, as is
 the case with the discussion in section 4 below.

\no
We start with the $\s$-model Lagrangian density
\be
{\cal L} = \frac{1}{2} Q_{IJ}\partial_+ X^I \partial_- X^I\ ,\qq Q_{IJ} = G_{IJ} + B_{IJ}\ ,
\ee
where $G$ and $B$ are the metric and the antisymmetric tensor in the NS sector, respectively.
We demand that the background fields are independent of some coordinate $X_0$ and denote the
rest of them by $X^i$. With the definitions\footnote{In our conventions $\s^\pm =\ha (\tau\pm \s)$.}
\be
J_+ = \frac{1}{2} Q_{i 0} \partial_+ X^i\ ,  \qq J_- = \frac{1}{2} Q_{ 0 i} \partial_- X^i\, ,
\qq V = - \frac{1}{2} Q_{ij} \partial_+ X^i \partial_- X^j\ ,
\ee
the momenta conjugate to $X_0$ is given by
\be
P_0 = \frac{\delta {\cal L}} {\delta \dot X_0} = G_{00} \dot X_0 + J_+ + J_-\  ,
\ee
and the Hamiltonian density obtained by performing the Legendre transform only on the active field $X_0$
is then
\ba
{\cal H} & = & {1\ov 2 G_{00}}P_0^2  +{G_{00}\ov 2} {X_0'}^2 -{1\ov G_{00}} P_0 (J_+ + J_-) + (J_+ - J_-)X_0'
\nonumber \\
&& +\ {1\ov 2 G_{00}} (J_+ + J_-)^2 + V \ .
\label{hami1}
\ea
The Poisson brackets between the conjugate phase space variables are
\be
\{ X_0(\s),P_0(\s')\}=\d(\s\!-\! \s')\ ,\qq \{ X_0(\s),X_0(\s')\} =  \{ P_0(\s),P_0(\s')\}=0 \ ,
\ee
where here and subsequently we suppress the $\tau$-dependence since we deal with equal time brackets.
Under the following transformation to a new set of phase space variables (which
preserves the above symplectic structure)
\be
P_0 = \tilde X_0^\prime \ ,  \quad X_0^\prime = \tilde{P_0} \ ,
\label{cantr1}
\ee
the Hamiltonian density becomes
\ba
{\cal H}_{\rm CT} & = & {1\ov 2 G_{00}}\tilde {X'_0}^{2}  +{G_{00}\ov 2} \tilde P_0^2 -{1\ov G_{00}}
\tilde X_0'  (J_+ + J_-) + (J_+ - J_-)\tilde P_0
\nonumber\\
&& + \ {1\ov 2 G_{00}} (J_+ + J_-)^2 + V \ .
\label{hami2}
\ea
The T-dual model has a Hamiltonian density $\tilde {\cal H}$ of the same form as that in \eqn{hami1} with $X_0$,
$P_0$, $G_{00}$, $J_\pm $ and $V$ replaced with the corresponding tilded quantities.
What is remarkable, and the crux of the issue, is that the dual Hamiltonian can be brought into exactly
the same form as the original Hamiltonian after a redefinition of the background fields. That is by demanding
\be
H_{\rm CT}=\int d\s\ {\cal H}_{\rm CT} = \int d\s\ \tilde {\cal H} = \tilde H\ ,
\ee
we obtain
\be
\tilde J_\pm = \mp {J_\pm \ov G_{00}}\ ,\qq \tilde V = V + 2 {J_+J_-\ov G_{00}}\ ,
\label{dkk1}
\ee
from which we easily recover the Buscher T-duality rules
\ba
&& \tilde G_{00} = {1\ov G_{00}}\ ,\qq \tilde Q_{i0}= -{Q_{i0}\ov G_{00}}\ ,\qq \tilde Q_{0i}= {Q_{0i}\ov G_{00}}
\ ,
\nonumber\\
&& \tilde Q_{ij} = Q_{ij} -{Q_{i0} Q_{0j}\ov G_{00}}\ .
\label{tdualbbo}
\ea
We mention that the transformation of worldsheet derivatives under the canonical transformation can be computed as
\be
\del_+ \tilde X_0 = G_{00}\del_+ X_0 + Q_{i 0} \del_+ X^i\ ,\qq
\del_- \tilde X_0 = - G_{00}\del_- X_0 + Q_{0 i} \del_- X^i\ .
\label{woordd}
\ee
Hence, the transformation of the differential involves the Hodge
dual on the worldsheet, i.e. $d\tilde X_0 = G_{00} \star d X_0 + \dots $.

\no
Finally, we note that the energy momentum tensor can be written with either set of
background fields and worldsheet derivatives, i.e.
\be
T_{\pm\pm} = \ha G_{IJ}\del_\pm X^I \del_\pm X^J = \ha \tilde G_{IJ}\del_\pm \tilde X^I
\del_\pm \tilde X^J = \tilde T_{\pm \pm}\ .
\ee
The proof relies on the transformation of the background fields and worldsheet derivatives,
 \eqn{tdualbbo} and \eqn{woordd}.
For completeness we also note that there
is an obvious generating function (of the first kind) for the canonical transformation
\be
{\cal F} = -\int d\sigma\ X_0^\prime \tilde X_0\ ,\qq
\Pi = \frac{\delta {\cal F} } {\delta X_0}  \ , \qq  \tilde{\Pi} = - \frac{\delta {\cal F} }
{\delta \tilde X_0}\ .
\label{canbos}
\ee

\section{Fermionic T-duality as a canonical transformation}\label{secfermt}

We now consider the fermionic T-duality proposed by Berkovits and Maldacena.
We begin by considering the Lagrangian density
\be
{\cal L } = \frac{1}{2} L_{MN} \partial_+ Z^M \partial_- Z^N\ ,\qq L_{MN}= G_{MN} + B_{MN}\ ,
\label{allk2}
\ee
where $Z^M = (X^I,\th^\a)$ are coordinates on a superspace so that the $\theta$ variables are
anti-commuting fermions and the superfields $G$ and $B$ obey graded symmetrisation rules
\begin{equation}
G_{MN}= (-)^{MN}G_{NM}\,  , \quad B_{MN}= -(-)^{MN}B_{NM}\ ,
\end{equation}
where $(-)^{MN}$ is equal to $+1$ unless both $M$ and $N$ are spinorial indices in which case it is equal
to $-1$.  The lowest components of these superfields, when the indices run over bosonic
coordinates, are the target space metric and B-field.

\no
We assume that the action is invariant under a shift symmetry in one of the fermionic directions $\th^1$
(which henceforth we will denote simply by $\theta$)
and that the background superfield is independent of this coordinate (this is much
the same as working in adapted coordinates for regular bosonic T-duality).  We define $Z^\m$ as
running over all bosonic and fermionic directions except $\theta$.
It is also helpful to define
\be
\label{fermJV}
 {\cal J}_+ = \ha L_{\m 1} \del_+ Z^\m \ ,\qq {\cal J}_- =-\ha (-1)^\m L_{1 \m} \del_- Z^\m\ ,
\qq {\cal V} = -\ha L_{\m\n} \del_+ Z^\m \del_- Z^\n\ .
\ee
Note that ${\cal J}_\pm $ are fermionic.
Then the $\s$-model Lagrangian can be written as
\begin{equation}
\label{fermL}
{\cal L } =  -B_{11} \dot{\theta} \th^\prime
+ ( \dot{\th} + \th^\prime) {\cal J}_{-} + {\cal J}_{+} ( \dot{\th} - \th^\prime) - {\cal V}  \ .
\end{equation}
The equation of motions from varying $\th$ is
\be
\d\th: \qq \dot B_{11} \th'-B_{11}' \dot\th + (\dot {\cal J}_+ - \dot {\cal J}_-)
 - ({\cal J}_+ + {\cal J}_-)' =0\ .
\label{eqmmot}
\ee
The canonical momenta conjugate to $\theta$ is given by
\be
\Pi = \frac{\delta {\cal L}} { \delta \dot{\theta}} = - B_{11} \th^\prime - {\cal J}_+ + {\cal J}_-\ ,
\label{defpi}
\ee
and the corresponding equal time Poisson Brackets are
\be
\{\th(\s), \Pi(\s^\prime) \} = - \delta(\s - \s^\prime)\ ,\qq \{\th(\s), \th(\s^\prime) \} =
\{\Pi(\s), \Pi(\s^\prime) \} =0 \ .
\label{fermioPB}
\ee
The sign convention in the first bracket
is a consequences of the fermionic nature of $\th$ and the fact that derivatives act from the left.

\subsection{The constrained system}

Since the Lagrangian is first order in time derivatives the velocities can not be solved in terms of momenta,
instead we have an anticommuting constraint
\be
f = \Pi  + B_{11} \th^\prime + {\cal J}_+ - {\cal J}_- \approx 0\ .
\label{constrr}
\ee
The naive Hamiltonian density is given as
\be
{\cal H} =\dot{\th} \Pi -  {\cal L} =  - \th^\prime  ({\cal J}_{+} + {\cal J}_{-})  + {\cal V}  \  ,
\ee
however, this should be amended to take account of the constraint.
We follow the Dirac procedure\footnote{See \cite{henneauxbook} for a detailed treatment of constrained dynamics.}  by first modifying the Hamiltonian with an,
as yet unknown, local function $\l(\tau,\s)$ which resembles an anticommuting Lagrange multiplier
\be
H_{\rm tot} = \int d\sigma \,  ({\cal H}   + \lambda f )\ .
\label{okww}
\ee
We now need to check whether any secondary constraints are produced by considering
the time evolution of the constraint and demanding that
\be
\dot{f}(\sigma) =  \{ f(\sigma) , H_{\rm tot} \} \approx 0\, .
\ee
In order to calculate this time evolution it is necessary to know
\ba
\{ f(\sigma) , f(\sigma^\prime) \} &=&  \{ \Pi(\s) + B_{11} \th^\prime(\s)  ,
\Pi(\s^\prime) +  B_{11}(\s^\prime) \th^\prime(\s^\prime)   \}
\nonumber \\
&=& B_{11}(\s^\prime)  \{ \Pi(\s) , \th^\prime(\s^\prime)   \} +   B_{11}(\s) \{  \th^\prime(\s)  , \Pi(\s^\prime)  \}
\nonumber \\
&=& (  B_{11}(\s^\prime)  - B_{11}(\s) ) \frac{ \partial}{\partial \sigma}  \delta( \sigma - \sigma^\prime)
\\
&=&   B_{11}^\prime ( \s^\prime)   \delta ( \s - \s^\prime)\ ,
\nonumber
\ea
where we have made use of the identifications $x \delta (x) = 0$ and
$x\d'(x)=-\d(x)$ (to be understood in a distributional sense).
Note that since the Poisson bracket of these constraints is non-zero
they are second class constraints; to consider the quantization
of the theory one should upgrade Poisson brackets to Dirac brackets.

\no
We also need that
\be
\{  f(\sigma) ,  {\cal H}(\s')    \} = - \{ \Pi , \theta^\prime ({\cal J}_+ + {\cal J}_-) \} \\
=({\cal J}_+ + {\cal J}_-)(\s') \frac{ \partial }{ \partial \s^\prime}  \delta ( \s - \s^\prime)  \, .
\ee
Then the time evolution is given by
\ba
\dot{f}(\sigma) &=&  \{ f(\sigma) , H_{\rm tot} \} =  \int d\sigma^\prime \, ({\cal J}_+ + {\cal J}_-)(\s')
   \frac{ \partial }{ \partial \s^\prime}  \delta ( \s - \s^\prime)       -
\lambda(\s^\prime)  B_{11}^\prime ( \s^\prime)   \delta ( \s - \s^\prime)\nonumber   \\
&=&   - ({\cal J}_+ + {\cal J}_-)^\prime       - \lambda(\s)  B_{11}^\prime ( \s)        \, ,
\ea
note that the minus sign in the second factor is due to the fact that $\lambda(\s)$ is anticommuting.
Demanding that $ \dot{f}(\sigma) \approx 0$ does not produce a new constraint but instead fixes the
Lagrange multiplier function as
\be
\lambda(\s) = - \frac{({\cal J}_+ + {\cal J}_-)^\prime  }{ B_{11}^\prime ( \s) }\ .
\label{solla}
\ee
Thus, the total Hamiltonian density is given by the integrand in \eqn{okww}
upon substituting \eqn{solla}. We obtain
\be
{\cal H}_{\rm tot} =    - \th^\prime  ({\cal J}_{+} + {\cal J}_{-})  + {\cal V}
 -  \frac{({\cal J}_+ + {\cal J}_-)^\prime  }{ B_{11}^\prime }  (\Pi  + B_{11} \th^\prime + {\cal J}_+ - {\cal J}_-  )\ .
\label{fhh1}
\ee
Having established the appropriate Hamiltonian for our constrained system let's verify that the time
evolution of $\th$ indeed gives rise to the equations of motion \eqn{eqmmot}.
We easily compute that
\be
\dot \th = \{\th,H_{\rm tot}\} = - {({\cal J}_+ + {\cal J}_-)'\ov B_{11}'}\
\label{ev1}
\ee
and that
\be
\dot\Pi = \{\Pi,H_{\rm tot}\}
= \Big({B_{11}\ov B_{11}'}({\cal J}_+ + {\cal J}_-)' - ({\cal J}_+ + {\cal J}_-)\Big)' =
 - (B_{11} \dot \th  + {\cal J}_+ + {\cal J}_-)'\ ,
\label{ev2}
\ee
where in the second equality we have used \eqn{ev1}. Then from the definition of $\Pi$ in \eqn{defpi}
the last equation becomes identical to \eqn{eqmmot}.

\subsection{The canonical transformation}

Consider now the transformation of phase-space variables
\be
\label{fermCT}
\th'=\tilde \Pi \ ,\qq \Pi =-\tilde \th'\ ,
\ee
which leaves invariant the Poisson brackets \eqn{fermioPB}. Under this transformation the Hamiltonian density
in \eqn{fhh1} becomes
\be
{\cal H}_{\rm tot, CT} =    - \tilde \Pi  ({\cal J}_{+} + {\cal J}_{-})  + {\cal V}
 -  \frac{({\cal J}_+ + {\cal J}_-)^\prime  }{ B_{11}^\prime }  (-\tilde \th'  + B_{11} \tilde \Pi + {\cal J}_+ - {\cal J}_-  )\ .
\label{fhh2}
\ee
As in the bosonic case, we would like to identify the Hamiltonian corresponding to this density
with that for the T-dual model. This means with a Hamiltonian whose density is as in \eqn{fhh1} with the
replacement of $\th$, $\Pi$, $B_{11}$, ${\cal J}_\pm$ and ${\cal V}$ with their tilded counterparts.
Comparing the coefficients of terms with $\tilde \Pi$ we obtain the condition
\be
\tilde \Pi :\quad - ({\cal J}_{+} + {\cal J}_{-}) + {B_{11}\ov B_{11}'} ({\cal J}_{+} + {\cal J}_{-})' =
{(\tilde{\cal J}_{+} + \tilde {\cal J}_{-})'\ov \tilde B_{11}'}\ ,
\label{pppp}
\ee
from which we deduce, after some algebraic manipulations, that
\be
\tilde B_{11}=-{1\ov B_{11}}\ ,\qq
\tilde{\cal J}_{+} + \tilde {\cal J}_{-} =  {{\cal J}_{+} + {\cal J}_{-}\ov  B_{11}}\ .
\label{b11j}
\ee
In addition, comparing the coefficients of terms with $\tilde \th'$ we obtain
\be
\tilde \th'  :\quad  -{({\cal J}_{+} + {\cal J}_{-})'\ov  B_{11}'} = - (\tilde {\cal J}_{+} + \tilde {\cal J}_{-})+
{\tilde B_{11}\ov \tilde B_{11}'} (\tilde {\cal J}_{+} + \tilde {\cal J}_{-})'\ ,
\ee
which is the tilded counterpart of \eqn{pppp} leading again to \eqn{b11j}.
Comparing the rest of the terms we obtain
\be
{\cal V} -{1\ov B_{11}'} ({\cal J}_{+} +  {\cal J}_{-})' ({\cal J}_{+} -  {\cal J}_{-})=
\tilde {\cal V} -{1\ov \tilde B_{11}'}
(\tilde{\cal J}_{+} + \tilde {\cal J}_{-})' (\tilde{\cal J}_{+} - \tilde {\cal J}_{-})\ ,
\ee
leading to the conditions
\be
\tilde{\cal J}_{+} - \tilde {\cal J}_{-} ={{\cal J}_{+} - {\cal J}_{-} \ov B_{11}}\ , \qq
\tilde{{\cal V}} = {\cal V} + 2 { {\cal J}_+ {\cal J}_-\ov B_{11}} \ .
\label{b22j}
\ee

\no
Combining \eqn{b11j} with \eqn{b22j} we obtain the Fermionic T-duality rules
\ba
&& \tilde B_{11} = -{1\ov B_{11}}\ ,\qq \tilde L_{\m 1}= {L_{\m 1}\ov B_{11}}\ ,\qq \tilde L_{1 \m}=
{L_{1\m}\ov B_{11}}
\ ,
\nonumber\\
&& \tilde L_{\m\n}= L_{\m\n} -{L_{1\n}L_{\m 1}\ov B_{11}}\ .
\label{tdualbbof}
\ea
Similarly to the bosonic case one may compute the transformation of the worldsheet derivatives of
$\th$ under the canonical transformation. We compute that
\be
\del_+\tilde \th = B_{11}  \del_+ \th + L_{\m 1}\del_+ Z^\m \ ,
\qq \del_-\tilde \th = B_{11}  \del_- \th - (-1)^\m L_{1\m}\del_- Z^\m\ .
\label{woorddf}
\ee
Hence, unlike the bosonic case the transforation of the differential does not involve the Hodge
dual on the worldsheet, i.e. $d\tilde \th = B_{11} d\th + \dots $.

\no
As in the bosonic case the energy momentum tensor can be written with either set of
background fields and worldsheet derivatives
\be
T_{\pm\pm} = \ha G_{MN}\del_\pm Z^M \del_\pm Z^M = \ha \tilde G_{MN}\del_\pm \tilde Z^M
\del_\pm \tilde Z^N = \tilde T_{\pm \pm}\ ,
\ee
where in the proof we have used \eqn{tdualbbof} and \eqn{woorddf}.
In addition,
the analog of the bosonic generating function \eqn{canbos} in the fermionic case is
\be
{\cal F} = \int d\sigma\ \theta^\prime \tilde{\theta}\ ,\qq
\Pi = \frac{\delta {\cal F} } {\delta \theta}  \ , \qq  \tilde{\Pi} = - \frac{\delta {\cal F} }
{\delta \tilde{\theta}}\ .
\ee

\no
Finally, we note that we could have followed a similar path leading to the fermionic
T-duality transformation rules by using the Dirac brackets for our canonical
variables which we include for
completeness
\ba
&& \{\th(\s_1),\th(\s_2)\}_D = -{\d(\s_1\!-\!\s_2)\ov B'_{11}(\s_2)}\ ,
\nonumber\\
&& \{\Pi(\s_1),\Pi(\s_2)\}_D = -\del_{\s_1}\del_{\s_2} \left({B_{11}^2(\s_1)\ov B_{11}'(\s_1)}
\d(\s_1\!-\!\s_2)\right)\ ,
\\
&& \{\th(\s_1),\Pi(\s_2)\}_D = -\d(\s_1\!-\!\s_2) -{B_{11}(\s_1)\ov B_{11}'(\s_1)} \d'(\s_1\!-\!\s_2)\ .
\nonumber
\ea
This procedure would have allowed us to set the constraint \eqn{constrr} strongly to zero in various expressions.
Then, in addition to equating the Hamiltonians, one should require that the constraint \eqn{constrr} is
actually preserved by the transformation.

\section{Canonical T-duality in the pure spinor formalism }

Many important string backgrounds have non zero RR fluxes, the most notable being $AdS_5\times S_5$.
 It is thus important to understand the action of T-duality on RR fields.   This was first established from
 a supergravity perspective \cite{Bergshoeff:1995as}  and later by means of a Buscher procedure in the
  Green Schwarz form of the superstring \cite{Kulik:2000nr,Cvetic:1999zs}.
  More recently the Buscher procedure was applied to the pure spinor form
  of the superstring \cite{Benichou:2008it,Chandia:2009yv}.

\subsection{A brief introduction and generalities}

The pure spinor approach to the superstring proposed by Berkovits
combines the virtues of the RNS formalism with the those of the GS formalism.
In particular, it allows one to describe the superstring in general
curved backgrounds with non-trivial Ramond--Ramond sectors.
We refer the reader to the original papers as well as the helpful reviews
on this subject for more of the details of the
formalism \cite{Berkovits:2007wz,Bedoya:2009np,Berkovits:2001ue} .

\no
The Lagrangian density in a curved background is given by
\begin{eqnarray}
\cL &=&
\frac{1}{2}L_{MN}(Z)\partial_+ Z^M \partial_- Z^N
+  P^{\alpha \hat{\beta}}(Z) d_{\alpha} \hat{d}_{\hat{\beta}} + E^{\alpha}_M (Z) d_{\alpha} \partial_- Z^M
\nonumber \\
  & & +\
E^{\hat{\alpha}}_M (Z) \hat{d}_{\hat{\alpha}} \partial_+ Z^M
+ \Omega_{M \alpha}\,^{\beta}(Z) \lambda^{\alpha} w_{\beta} \partial_- Z^M +
\hat{\Omega}_{M \hat{\alpha}}\,^{\hat{\beta}}(Z) \hat{\lambda}^{\hat{\alpha}}
\hat{w}_{\hat{\beta}} \partial_+ Z^M
\label{action}\\
  & &  +\ C_{\alpha}^{\beta \hat{\gamma}}(Z) \lambda^{\alpha} w_{\beta}
  \hat{d}_{\hat{\gamma}} + \hat{C}_{\hat{\alpha}}^{\hat{\beta} \gamma}(Z)
  \hat{\lambda}^{\hat{\alpha}} \hat{w}_{\hat{\beta}} d_{\gamma}
+ S_{\alpha \hat{\gamma}}^{\beta \hat{\delta}}(Z) \lambda^{\alpha} w_{\beta}
  \hat{\lambda}^{\hat{\gamma}} \hat{w}_{\hat{\delta}}+ \cL_{\lambda} +
    \hat{\cL}_{\hat{\lambda}}\ .
\nonumber
\end{eqnarray}
In this action the fields $Z^M$ describe a mapping of the worldsheet into a superspace
$\mathbb{R}^{10|32}$ and can be broken up into a bosonic part and fermionic parts
$Z^M = (Z^m, \theta^\alpha , \hat{\theta}^{\hat{\alpha}})$.
In the type-IIA theory $\theta$ and $\hat{\theta}$ have opposing
chiralities whereas in the type-IIB theory they have the same chirality.
The remaining fields $\omega_\a$ and $\lambda^\a$  (and their hatted counterparts) are conjugate variables
and are bosonic spinor ghosts with kinetic terms  $\cL_{\lambda}$.
 $\lambda^\a$  obey the pure spinor constraints  $\lambda^\a \gamma^m_{\a \b} \lambda^\beta = 0$
 so that their contribution to the central charge cancels that coming from the $Z^M$.

\no
The fermionic field $d_\a$ is vital in the pure spinor construction since it is used in forming the BRST
operator
\begin{equation}
Q = \oint \lambda^\a d_\a \ ,\qq \hat Q =  \oint \hat \lambda^{\hat \a} \hat d_{\hat \a}\ .
\end{equation}
In flat space the nilpotency of this BRST operator is shown using the OPE of the $d_\a$
and also the pure spinor constraint.  For the curved space theory defined above demanding that
$Q$ is nilpotent and holomorphic constrains the background fields to obey the equations of motion of type-II supergravity.

\no
Other than the inclusion of a  Fradkin--Tseytlin term, the action \eqn{action}  represents
the most  general $\sigma$-model coupled to background fields whose interpretation we now summarise.
The superfield $L_{MN}(Z)$ is defined as in \eqn{allk2} and contains the metric and NS two-form.
The field $P^{\a \hat{\b}}$ contains the RR field strengths and has lowest component
\ba
&& P^{\a \hat{\b}}\big|_{\th,\hat \th=0} = -{ i \over 4}  e^{\phi} F^{\a\hat{b}}
\nonumber\\
&&\phantom{xx} =
   -{ i \over 4}  e^{\phi}
  \Big( (\gamma^m)^{\alpha \hat \beta} F_m + { 1 \over 3 !}
(\gamma^{ m_1 m_2 m_3})^{\alpha \hat \beta} F_{m_1 m_2 m_3} + { 1 \over 2} { 1 \over 5!}
 (\gamma^{ m_1\cdots m_5})^{\alpha \hat \beta} F_{m_1 \cdots  m_5}\Big),
\label{paab}
\ea
with a similar expression for the type-IIA theory involving even forms.
The field $E^{\alpha}_M$
is part of the super-vielbein, $\hat{\Omega}_{\mu \hat{\alpha}}{}^{\hat{\beta}}$
contains the (torsionfull) spin connection and  $S_{\alpha \hat{\gamma}}^{\beta \hat{\delta}}$
contains curvature terms.

\subsection{Bosonic T-duality}

We demand that the background fields entering \eqn{action} are independent of some bosonic
coordinate $X_0$ and denote the rest of them by $Z^\m$.
The action \eqn{action} corresponds to a Hamiltonian with density of the form \eqn{hami1}
but with $J_\pm$ and $V$ defined by
\ba
 && J_+ =  \ha L_{\m 0} \del_+ Z^\m + E^\a_0 d_\a + \Om_{0\a}{}^\b \l^\a \om_\b \ ,
\nonumber\\
&& J_-  =  \ha L_{0 \m } \del_- Z^\m + E^{\hat \a}_0 \hat d_{\hat \a} + \hat \Om_{0\hat\a}{}^{\hat \b}
\hat \l^{\hat \a} \hat \om_{\hat \b} \
\ea
and
\ba
\label{purespV}
 V & = & -\ha L_{\m\n}\del_+ Z^\m \del_- Z^\n - P^{\a\hat \b} d_{\a} \hat d_{\hat \b}
- E^\a_\m d_\a \del_- Z^\m  - E^{\hat \a}_\m \hat d_{\hat \a} \del_+Z^\m
\nonumber\\
&& \ - \Om_{\m\a}{}^\b \l^\a \om_\b \del_- Z^\m - \hat \Om_{\m\hat \a}{}^{\hat \b} \hat \l^{\hat \a}
 \hat \om_{\hat \b} \del_+ Z^\m
\\
&& -C_\a^{\b\hat \g} \l^\a \om_\b \hat d_{\hat \g} - \hat C_{\hat a}^{\hat \b \g}\hat \l^{\hat \a}
\hat\om_{\hat \b}  d_\g
 -S^{\b \hat \d}_{\a\hat \g} \l^\a \om_\b \hat \l^{\hat \g} \hat \om_{\hat \d} - \cL_\l - \hat \cL_{\hat \l}\ .
\nonumber
\ea
In this case the transformation \eqn{cantr1} should be accompanied with a transformations that changes the
chirality of the spinors. We may choose to change the chirality of either component corresponding to the hatted or
unhatted symbols. We choose to do so for the hatted ones, which implies that all hatted fermions
transform as
\be
\tilde {\hat \psi} = \G \hat \psi\ ,\qq \hat \psi = (\hat \th, \hat d, \hat \l, \hat \om)\ ,
\label{dkk12}
\ee
where $\gamma_1 \equiv \G $ is the gamma-matrix in the direction of the isometry. This transformation
clearly leaves invariant the Poisson brackets
between the spinorial fields due to the fact that $\G^2 = \mathbb{I}$.

\no
Using the first of \eqn{dkk1} and \eqn{dkk12} we obtain that
\ba
&& \tilde G_{00} = {1\ov G_{00}}\ ,\qq
\tilde L_{\m 0} = -{L_{\m 0}\ov G_{00}} \ ,\qq \tilde L_{0 \m } = {L_{0 \m }\ov G_{00}}\ ,
\nonumber\\
&& \tilde E^\a_0 = -G_{00}^{-1} E^\a_0\ ,\qq \tilde E^{\hat \a}_0 = G_{00}^{-1}  E^{\hat \b}_0
 \G_{\hat \b}{}^{\hat \a} \ ,
\\
&& \tilde \Om_{0\a}{}^\b = -G_{00}^{-1} \Om_{0\a}{}^\b\ ,\qq \tilde {\hat \Om}_{0\hat \a}{}^{\hat \b} =
G_{00}^{-1} \hat \Om_{0\hat \g}{}^{\hat \d} \G^{\hat \g}_{\hat \a} \G_{\hat \d}{}^{\hat \b}\ .
\nonumber
\ea
\no
Using the second of \eqn{dkk1} and \eqn{dkk12} as well as introducing the matrices
\be
A_M{}^N = \left(
            \begin{array}{cc}
              -G^{-1}_{00} & 0 \\
              -G_{00}^{-1} L_{0\m} & \d_\m{}^\n \\
            \end{array}
          \right)\ ,\qq
\hat A_M{}^N = \left(
            \begin{array}{cc}
              G^{-1}_{00} & 0 \\
              -G_{00}^{-1} L_{\m 0} & \d_\m{}^\n \\
            \end{array}
          \right)\ ,
\ee
we obtain that
\ba
&& \tilde L_{\m\n} = L_{\m\n}-{L_{0\n} L_{\m 0}\ov G_{00}}\ ,
\nonumber\\
&& \tilde E^\a_M =A_M{}^N E^\a_N \ ,\qq \tilde E^{\hat \a}_M =\hat A_M{}^N E^{\hat \b}_N \G_{\hat \b}{}^{\hat \a}\ ,
\\
&& \tilde \Om_{M \a}{}^\b= A_M{}^N \Om_{N \a}{}^\b\ ,\qq
\tilde {\hat \Om}_{M \hat \a}{}^{\hat \b
}= \hat A_M{}^N \hat \Om_{N \hat \g}{}^{\hat \d} \G_{\hat \d}{}^{\hat \b} \G^{\hat \g}{}_{\hat \a}\
\nonumber
\ea
and
\ba
&& {\tilde P}^{\a\hat \b} = \Big(P^{\a\hat \g}+ {2\ov G_{00}} E^\a_0 E^{\hat \g}_0 \Big)\G_{\hat \g}{}^{\hat \b}\ ,
\nonumber\\
&& \tilde C^{\a\hat \g}_\b = \Big(C^{\a\hat \g}_\b - {2\ov G_{00}} \Om_{0 \b}{}^\a E^{\hat \d}_0 \Big)
\G_{\hat \d}{}^{\hat \g}\ ,
\quad
\tilde {\hat C}^{\hat \a  \g}_{\hat \b} = \Big(C^{{\hat \d} \g}_{\hat \g}
- {2\ov G_{00}} E^{\g}_0  \hat \Om_{0 \hat \g}{}^{\hat \d} \Big )\G_{\hat \d}{}^{\hat \a} \G^{\hat \g}{}_{\hat \b}\ ,
\\
&& \tilde S^{\a\hat  \g}_{\b\hat \d}= \Big(S^{\a\hat \e}_{\b\hat \zeta} -{2\ov G_{00}} \Om_{0\b}{}^\a
\hat \Om_{0\hat \zeta}{}^{\hat \e}\Big )\G_{\hat \e}{}^{\hat \g} \G^{\hat \zeta}{}_{\hat \delta}\ .
\nonumber
\ea
In practice, the expression \eqn{paab} can be used to
read off the transformation rules of the various forms in the theory. The fact that we have changed the
chirality of the hatted fermions according to \eqn{dkk12} has the consequence that the bosonic
T-duality takes one from the type-IIA to the type-IIB and vice versa.

\subsection{Fermionic T-duality}
The treatment of fermionic T-duality in the pure spinor superstring follows exactly
the steps described in section \ref{secfermt}.  The Lagrangian is of the same form
as \eqn{fermL} but with ${\cal J}_\pm$ in \eqn{fermJV} replaced by
\ba
\label{purespJ}
{\cal J}_+ &=& \frac{1}{2}L_{\m 1}
\partial_+ Z^\mu + E_1^\a d_\a + \Om_{1\a}{}^\b \lambda^\a \omega_\beta\, ,  \nonumber \\
{\cal J}_- &=& - (-1)^\mu\frac{1}{2} L_{1 \mu} \partial_- Z^\mu  - E_1^{\hat{\a}} \hat{d}_{\hat{\a}}
- \hat{\Om}_{1 \hat{\a}}{}^{\hat{\b}} \hat{\lambda}^{\hat{\a}} \hat{\omega}_{\hat{\beta}}
\ea
and the potential ${\cal V}$ given by the lengthy expression in \eqn{purespV}
but with the understanding that $Z^\mu$ runs over all coordinates except $\th$,
the fermionic direction along which we are dualising.

\no
We have seen, that under the canonical transformation \eqn{fermCT}, the transformed
\emph{total} Hamiltonian can be viewed as the Hamiltonian of a dual $\s$-model with the identifications
\be
\tilde{B}_{11}= - {1\ov B_{11}}\ , \qq \tilde{{\cal J}}_\pm = { {\cal J}_\pm\ov B_{11}} \ , 
\qq  \tilde{{\cal V}} = {\cal V} + 2 {{\cal J}_+ {\cal J}_-\ov B_{11}} \ .
\ee
Inserting the form of the currents \eqn{purespJ} and potential \eqn{purespV} into these
 relations yields the
 fermionic T-duality rules\footnote{Note that some of signs in the expressions presented here
 differ from those in equation 2.20 of \cite{Berkovits:2008ic}. In addition we have corrected a small typographical error in the transformation of $S^{\a\hat  \g}_{\b\hat \d}$.}
\ba
&& \tilde{B}_{11}= - {1\ov B_{11}}\ , \qq \tilde{L}_{\m 1} =  { L_{\m 1}\ov B_{11}} \ , 
\qq  \tilde{L}_{1 \m } = {L_{1 \m }\ov B_{11}}\  ,
 \nonumber \\
&& \tilde{L}_{\m \n} = L_{\m \n} -  {L_{1 \n} L_{\m 1} \ov B_{11}}\  ,
 \\
&& \tilde{P}^{\a \hat{\beta}} = P^{\a \hat{\b}} + 2{ E_1^\a E_1^{\hat{\beta}}\ov B_{11}} \, ,
\nonumber
\ea
\ba
&& \tilde{E}_M^\a =B_M{}^N E^\a_N\ ,\qq \tilde{E}_M^{\hat \a} =\hat B_M{}^N E^{\hat \a}_N\ ,
\nonumber\\
&& \tilde{\Om}_{M\a}{}^\b  =B_M{}^N  \tilde{\Om}_{N\a}{}^\b \ ,\qq
  \tilde{\hat \Om}_{M\hat \a}{}^{\hat \b}  =\hat B_M{}^N  {\hat \Om}_{N\hat \a}{}^{\hat \b}\ ,
\ea
where
\be
B_M{}^N = \left(
            \begin{array}{cc}
              B^{-1}_{11} & 0 \\
              -B_{11}^{-1} L_{1\m} & \d_\m{}^\n \\
            \end{array}
          \right)\ ,\qq
\hat B_M{}^N = \left(
            \begin{array}{cc}
              B^{-1}_{11} & 0 \\
              (-1)^\m B_{11}^{-1} L_{\m 1} & \d_\m{}^\n \\
            \end{array}
          \right)\
\ee
and
\ba
&& \tilde C^{\a\hat \g}_\b = C^{\a\hat \g}_\b + 2 B_{11}^{-1}   E_1^{\hat{\g}} \Om_{1 \b}{}^\a\, ,
\quad
\tilde{\hat {C}}^{\hat{\a} \g}_{\hat{\b}}  =  \hat {C}^{\hat{\a} \g}_{\hat{\b}}
  -  2 B_{11}^{-1} \Om_{1 \hat{\b}}{}^{\hat{\a}}E_1^\g  \,
\nonumber \\
&& \tilde{S}^{\a\hat  \g}_{\b\hat \d}=  S^{\a\hat  \g}_{\b\hat \d} + 2 B_{11}^{-1}
 \Om_{1 \b}{}^{\a} \hat{\Om}_{1 \hat{\d}} {}^{\hat{\g}} \, .
\ea

\vskip .3 in
\centerline{ \bf Acknowledgments}

\no
We would like to thank Ilya Bakhmatov and David Berman for helpful discussions. D.C.T thanks the University of
Patras for hospitality during a visit in which this work was initiated.
K. Siampos acknowledges support by the  Greek State Scholarship Foundation
(IKY) and D.C.T. is supported by a STFC studentship.


\bibliographystyle{JHEP}
\bibliography{DansBib}
\end{document}